\documentclass{moriond}

\def\be{\begin{equation}}
\def\ee{\end{equation}}
\def\bea{\begin{eqnarray}}
\def\eea{\end{eqnarray}}

\usepackage{amsmath}	
\usepackage{amssymb}	
\usepackage{mathtools}

\begin{document}
\vspace*{4cm}
\title{\MakeUppercase{Axisymmetric equilibrium models for magnetised neutron stars in scalar-tensor theories}}

\author{\MakeUppercase{J. Soldateschi$^{1,2,3}$, N. Bucciantini$^{2,1,3}$, and L. Del Zanna$^{1,2,3}$}}

\address{$^{1}$Dipartimento di Fisica e Astronomia, Università degli Studi di Firenze, Via G. Sansone 1, 50019 Sesto F. no (Firenze), Italy \\
$^{2}$INAF – Osservatorio Astrofisico di Arcetri, Largo E. Fermi 5, 50125 Firenze, Italy \\
$^{3}$INFN – Sezione di Firenze, Via G. Sansone 1, 50019 Sesto F. no (Firenze), Italy}

\maketitle\abstracts{
General relativity (GR) probably is not the definitive theory of gravity, due a number or issues, both from the theoretical and from the observational point of view. Alternative theories of gravity were conceived to extend GR and account for such issues. Among the most promising ones are scalar-tensor theories (STTs), which predict an enrichment of the phenomenology of compact objects, like neutron stars (NSs). We updated the well-tested \texttt{XNS} code to numerically solve the Einstein-Maxwell equations for a stationary, magnetised NS in a class of STTs containing the spontaneous scalarisation phenomenon. We found that there exist ``quasi-universal relations'' among the mass, radius, scalar charge and magnetic deformation of a NS that are true independently of the equation of state (EoS), both in GR and in STTs. This result could potentially provide new tools to test STTs and the magnetic field geometry inside NSs.
}
 
\section{Introduction}\label{sec:introduction}
While GR has collected an incredible number of successes, there are some issues that seem to jeopardise its validity. In fact there are several arguments that seem to imply that GR is not the definitive theory of gravity. From the theoretical point of view, the Einstein-Hilbert action, which yields Einstein's equations of GR, is not the only possible action describing the gravitational interaction; in fact, there exist infinitely-many alternatives to GR~\cite{capozziello_2011,papantonopoulos_2015}. Moreover, a consistent quantum theory of gravity GR does not yet exist~\cite{capozziello_2011}. From the observational side, it is well known that we lack an explanation for the ``dark sector''; while one solution is to introduce an unknown form of matter-energy filling most of the Universe content, a different approach is to introduce modifications to GR to account for this problem. For these reasons, many theories of gravity alternative to GR have been developed~\cite{capozziello_2011}. Arguably, the most studied and promising ones are STTs, because they are the most simple extensions of GR~\cite{papantonopoulos_2015}, are predicted to be the low-energy limit of some possible theories of quantum gravity~\cite{damour_2002}, most of them respect the well-tested weak equivalence principle, and they seem to be free of pathologies common to other alternatives to GR~\cite{defelice_2006}.
The guiding principle of STTs is that of adding a scalar field to the gravitational action, in a way as to be non-minimally coupled to the metric. Some of these theories predict the existence of a non-perturbative strong field effect called ``spontaneous scalarisation''~\cite{damour_1993}, which allows the scalar field to exponentially grow in magnitude inside compact material objects, i.e. neutron stars. The importance of this phenomenon is that of allowing scalarised NSs to develop potentially observable modifications to GR, while still fulfilling the tight observational constraints in the weak-field regime~\cite{shao_2017}.
In our work, we studied for the first time the scenario of magnetised models of static, axisymmetric NSs in a class of massless STTs containing the spontaneous scalarisation phenomenon. We extended the well-tested \texttt{XNS} code~\cite{pili_2014,pili_2017}, which is based on the XCFC approximation for the metric, to the case of a generic STT and for any tabulated EoS~\cite{soldateschi_2020,soldateschi_2021}. We computed models of NSs, both in GR and in STT, with a variety of EoSs (detailed in Sec.~\ref{sec:eos}) and then used a ``principal component analysis'' (PCA) algorithm to find correlations among the main potentially observable quantities, allowing us to find some EoS-independent relations between them (see Sec.~\ref{sec:results} for the details).

\section{Scalar-tensor theories in a nutshell}\label{sec:stts}
In the `Bergmann-Wagoner formulation'~\cite{bergmann_1968,wagoner_1970} of STTs, the action in the Jordan frame (J-frame) is
\begin{equation}
\renewcommand{\arraystretch}{1.2}
\begin{array}{rc@{\,}c@{\,}l}
	\tilde{S}&= \frac{1}{16\pi}\int \mathrm{d}^4x \sqrt{-\tilde{g}}\left[ \varphi \tilde{R} - \frac{\omega (\varphi)}{\varphi} \tilde{\nabla} _\mu \varphi \tilde{\nabla} ^\mu \varphi - U(\varphi) \right] + \tilde{S}_\mathrm{p}\left[ \tilde{\Psi} , \tilde{g}_{\mu \nu} \right]  \; ,

\end{array}
\label{eq:joract}
\end{equation}
where $\tilde{g}$ is the determinant of the spacetime metric $\tilde{g}_{\mu \nu}$, $\tilde{\nabla} _\mu$ its associated covariant derivative, $\tilde{R}$ its Ricci scalar, while $\omega (\varphi)$ and $U(\varphi)$ are, respectively, the coupling function and the potential of the scalar field $\varphi$, and $\tilde{S}_\mathrm{p}$ is the action of the physical fields $\tilde{\Psi}$. Quantities denoted with a tilde are computed in the J-frame. In the Einstein frame (E-frame), the action is obtained by making the conformal transformation $\bar{g}_{\mu \nu}= \mathcal{A}^{-2}(\chi) \tilde{g}_{\mu \nu}$, where $\mathcal{A}^{-2}(\chi) =\varphi (\chi)$ and $\chi$ is a redefinition of the scalar field in the E-frame, related to $\varphi$ by $\mathrm d \chi/\mathrm d\ln \varphi = \sqrt{[\omega(\varphi) +3]/4}$. Quantities denoted with a bar are computed in the E-frame. In the case of a massless scalar field, which we adopt, $U(\varphi)=0$. In the E-frame the scalar field is minimally coupled to the metric, thus Einstein's field equations retain the same form as in GR in this frame, keeping into account the fact the energy-momentum tensor is now the sum of the physical one and of the scalar field one. Instead, the scalar field is minimally coupled to the physical fields in the J-frame, thus MHD equations in this frame have the same expression as in GR. In STTs we have an additional equation to solve for the scalar field. In the Einstein frame it reads
\begin{equation}\label{eq:scal}
	\bar{\nabla} _\mu \bar{\nabla} ^\mu\chi = -4\pi \alpha _\mathrm{s} \bar{T}_{\mathrm{p}} \; ,
\end{equation}
where $\bar{\nabla} _\mu$ is the covariant derivative associated to the E-frame metric $\bar{g}_{\mu \nu}$, $\bar{T}_{\mathrm{p}}= \bar{g}_{\mu \nu}\bar{T}_{\mathrm{p}}^{\mu \nu}$, $\bar{T}^{\mu \nu}_{\mathrm{p}}$ is the physical energy-momentum tensor in the E-frame and $\alpha _\mathrm{s}(\chi)~=~d \ln \mathcal{A}(\chi) / d\chi $. We adopted an exponential coupling function~\cite{damour_1993} $\mathcal{A}(\chi)~=~\exp \{ \alpha _0 \chi + \beta _0 \chi ^2 /2 \} $. The $\alpha _0$ parameter regulates the weak-field effects, while the $\beta _0$ parameter controls spontaneous scalarisation.

\section{A selection of equations of state}\label{sec:eos}
We performed our simulations using 13 different EoSs, chosen to span different calculation methods and particle contents, ranging from zero-temperature, $\beta$-equilibrium, purely nucleonics ones, to EoSs containing other particles and computed at a finite temperature. We also considered strange quark matter EoSs and a polytropic one. All EoSs were chosen to reach a maximum mass of $~2.05$M$_\odot$, have a radius of $~10-14$km for $1.4$M$_\odot$ mass models and satisfy various nuclear physics~\cite{fortin_2016} and stiffness~\cite{guerra_2019} constraints. We considered the following EoSs: APR~\cite{typel_2013}, SLY9~\cite{typel_2013}, BL2~\cite{typel_2013} (which is named ``BLEOS with crust'' in the CompOSE database), DDME2~\cite{fortin_2016}, NL3$\omega \rho$~\cite{fortin_2016}, SFHo~\cite{typel_2013}, DDME2-Y~\cite{fortin_2016} (equivalent to DDME2 with the addition of hyperons), NL3$\omega \rho$-Y~\cite{fortin_2016} (equivalent to NL3$\omega \rho$ with the addition of hyperons), BH8~\cite{typel_2013} (which is named ``QHC18'' in the CompOSE database), BF9~\cite{typel_2013} (which is named ``QHC19-B'' in the CompOSE database), SQM1~\cite{alcock_1986}, SQM2~\cite{fraga_2014}, POL2~\cite{bocquet_1995}.

\section{Quasi-universal relations}\label{sec:results}
We focused on magnetised, stable models of static NSs with maximum magnetic fields of $B_\mathrm{max}~\lesssim ~10^{17}$G: in this regime the quadrupolar deformation of the star $e=(I_{zz}-I_{xx})/I_{zz}$, where $I_{xx}$ and $I_{zz}$ are the Newtonian moments of inertia, is well approximated by the perturbative formulas
\begin{equation}\label{eq:distcoeff}
    |e| = c_\mathrm{B} B^2_\mathrm{max} + \mathcal{O}\left( B^4_\mathrm{max} \right) ,\;
    |e| = c_\mathrm{H} \frac{\mathcal{H}}{W} + \mathcal{O}\left( \frac{\mathcal{H}^2}{W^2} \right) ,\;
    |e| = c_\mathrm{s} B^2_\mathrm{s} + \mathcal{O}\left( B^4_\mathrm{s} \right) ,\;
\end{equation}
where $c_\mathrm{B}$, $c_\mathrm{H}$ and $c_\mathrm{s}$ are called the ``distortion coefficients'', $B_\mathrm{max}$ is normalised to $10^{18}$G, $\mathcal{H}$ is the magnetic energy~\cite{soldateschi_2021}, $W$ is the gravitational binding energy~\cite{soldateschi_2020} and $B_\mathrm{s}$ is the magnetic field at the pole. We computed the distortion coefficients for several models, both in GR and in STT with $\beta _0 \in \{ -6,-5.75,-5.5,-5 \}$, using the EoSs listed in Sec.~\ref{sec:eos}, in the case of purely toroidal and purely poloidal magnetic fields (except for $c_\mathrm{s}$, which is defined only in the purely poloidal case). Moreover, we performed a PCA to find EoS-independent relations, which we call ``quasi-universal relations'', between the distortion coefficients and the main observables describing our models: the Komar mass $M_\mathrm{k}$, the circumferential radius $R_\mathrm{c}$ and the scalar charge $Q_\mathrm{s}$~\cite{soldateschi_2020}. In GR we found that the following formulas approximate $c_\mathrm{B}$, $c_\mathrm{H}$, $c_\mathrm{s}$, with a relative error $|c^\mathrm{PCA}_\mathrm{B,H,s}-c_\mathrm{B,H,s}|/c_\mathrm{B,H,s}$ mostly under $\sim 15\%, 2\%, 5\%$ respectively, for all EoSs except SQM1, SQM2 and POL2:
\begin{equation}\label{eq:cb_pca_gr}
        c^\mathrm{PCA}_\mathrm{B} = 
            \begin{cases} 
                0.13 R_{10}^{5.45} M_{1.6}^{-2.41} \;{\rm for\; poloidal}\\
                0.25 R_{10}^{5.03} M_{1.6}^{-2.07} \;{\rm for\; toroidal}
            \end{cases},
\end{equation}
\begin{align}\label{eq:ch_pca_gr}
        c^\mathrm{PCA}_\mathrm{H} = 
            \begin{cases}
                5.77 - 0.77 R_{10} - 4.14 M_{1.6} - 0.27 M_{1.6}^2 + 0.07 R_{10}^2 + 2.28 M_{1.6} R_{10} \;{\rm for\; poloidal}\\
                7.02 - 5.22 R_{10} - 2.76 M_{1.6} - 0.12 M_{1.6}^2 + 1.92 R_{10}^2 + 1.51 M_{1.6} R_{10} \;{\rm for\; toroidal}
            \end{cases},
\end{align}
\begin{equation}\label{eq:cs_pca_gr}
    \begin{split}
        c^\mathrm{PCA}_\mathrm{s} &= 2.97 R_{10}^{4.61} M_{1.6}^{-2.80} \; ,
    \end{split}
\end{equation}
where $R_{10} = R_\mathrm{c}/10\mathrm{km}$ and $M_{1.6} = M_\mathrm{k}/1.6$M$_\odot$. The relative error when using these formulas for NSs described by the POL2 EoS reaches $\sim 90\%$ for $c_\mathrm{B}$ both in the poloidal and toroidal case, $\sim 7\% (\sim 20\%)$ for $c_\mathrm{H}$ in the poloidal (toroidal) case, $\sim 20\%$ for $c_\mathrm{s}$. If used in the case of SQM1 and SQM2, the relative error reaches $\sim 60\% (\sim 50\%)$ for for $c_\mathrm{B}$ for purely poloidal (toroidal) magnetic fields, $\sim 8\% (\sim 40\%)$ for $c_\mathrm{H}$ in the poloidal (toroidal) case, $\sim 40\%$ for $c_\mathrm{s}$.

In the case of STTs, we focused on finding an approximation for $\Delta c_\mathrm{B,H,s}~=~|c_\mathrm{B,H,s}~-~c^\mathrm{GR}_\mathrm{B,H,s}|$,  where $c^\mathrm{GR}_\mathrm{B,H,s}$ are the quasi-universal relations found in the GR case: Eq.s~\ref{eq:cb_pca_gr}-\ref{eq:ch_pca_gr}-\ref{eq:cs_pca_gr}. We found that the following formulas approximate $\Delta c_\mathrm{B}, \Delta c_\mathrm{H}, \Delta c_\mathrm{s}$, with a relative error mostly under $\sim~50\%, 7\%, 10\%$ respectively, for all EoSs except SQM1, SQM2 and POL2 and for any $\beta _0 \in \{ -6,-5.75,-5.5,-5 \}$:
\begin{equation}\label{eq:cb_pca_stt}
        \Delta c^\mathrm{PCA}_\mathrm{B} = 
            \begin{cases} 
                0.03 R_{10}^{8.23} M_{1.6}^{-5.08} Q_{1}^{2.60} \;{\rm for\; poloidal}\\
                0.06 R_{10}^{5.96} M_{1.6}^{-3.52} Q_{1}^{1.95} \;{\rm for\; toroidal}
            \end{cases},
\end{equation} 
\begin{equation}\label{eq:ch_pca_stt}
        \Delta c^\mathrm{PCA}_\mathrm{H} = 
            \begin{cases} 
                1.96 R_{10}^{0.72} M_{1.6}^{-1.96} Q_{1}^{1.54} \;{\rm for\; poloidal}\\
                1.49 R_{10}^{0.75} M_{1.6}^{-1.81} Q_{1}^{1.55} \;{\rm for\; toroidal}
            \end{cases},
\end{equation}
\begin{equation}\label{eq:cs_pca_stt}
        \Delta c^\mathrm{PCA}_\mathrm{s} = 
                0.92 R_{10}^{4.77} M_{1.6}^{-4.50} Q_{1}^{1.71} ,
\end{equation}
where $Q_{1}$ is $Q_\mathrm{s}$ normalised to 1M$_\odot$. The relative error when using these formulas for NSs described by the POL2 EoS remains roughly unaffected for $\Delta c_\mathrm{B}$, but reaches $\sim 15\% (\sim 40\%)$ for $\Delta c_\mathrm{H}$ in the poloidal (toroidal) case and $\sim 80\%$ for $\Delta c_\mathrm{s}$. If used in the case of SQM1 and SQM2, the relative error remains mostly under $\sim 100\%$ for $\Delta c_\mathrm{B}$ for both purely poloidal and toroidal magnetic fields. Instead, it reaches $\sim 40\% (\sim 50\%)$ for $\Delta c_\mathrm{H}$ in the poloidal (toroidal) case and $\sim 70\%$ for $\Delta c_\mathrm{s}$.
\\\\
The quasi-universal relations we found may be useful in shedding light into some of the major uncertaintes in NS astrophysics. For example, on the one hand $c_\mathrm{s}$ can be obtained from its definition in Eq.~\ref{eq:distcoeff} if both $B_\mathrm{s}$ and $e$ are known. The latter can be obtained by observing the emission of continuous gravitational waves from a NS, given that its strain is $h_0 \propto eI$ and the moment of inertia $I$ of the star around its rotation axis can be obtained in an EoS-independent way~\cite{breu_2016} by knowing its mass and radius. On the other hand, Eq.~\ref{eq:cs_pca_gr} can be applied. Depending on whether $c_\mathrm{s}<c^\mathrm{PCA}_\mathrm{s}$ or $c_\mathrm{s}>c^\mathrm{PCA}_\mathrm{s}$ some conclusions can be drawn: in the first case, a toroidal component must also be present, which counteracts the deformation caused by the poloidal component by reducing it; in the second case, another source of deformation, other than the magnetic field, must be present (because a purely poloidal field is an extremal magnetic configuration, thus causing the maximum possible magnetic deformation of the star). If, instead, $I$ is not computed in an EoS-independent way, one can use the comparison between $c_\mathrm{s}$ and $c^\mathrm{PCA}_\mathrm{s}$ to draw some conclusions about the EoS: if $c_\mathrm{s}>c^\mathrm{PCA}_\mathrm{s}$ either there is another source of deformation or the EoS predicts a moment of inertia that is not compatible with the observed deformation of the star, and is thus not consistent. Similar conclusion can be drawn for $c_\mathrm{B}$ and $c_\mathrm{H}$. Moreover, the quasi-universal relations we found can be useful to quickly determine the magnetic deformation of a NS model without going through a full numerical simulation. Finally, in the case of STTs, the scalar charge is also unknown. In this case, relations Eq.s~\ref{eq:cb_pca_stt}-\ref{eq:ch_pca_stt}-\ref{eq:cs_pca_stt} can be used to determine whether an observed NS deformation is compatible with a non-zero scalar charge.

%
%
%


%

\section*{Acknowledgments}

The authors acknowledge financial support from the INFN Teongrav collaboration.
%
%

\bibliography{soldateschi}
\end{document}